\documentclass[10pt, conference, compsocconf]{IEEEtran}
\usepackage{graphicx,url,multirow,subfigure,caption,listings,color,
            enumerate,breqn,flushend,array}
\definecolor{googleBlue}{rgb}{0.1490196,0.3019608,0.7529412}
\definecolor{googleRed}{rgb}{0.8196078,0.1294118,0.05882353}
\definecolor{googleYellow}{rgb}{0.9921569,0.5176471,0.03137255}
\definecolor{googleGreen}{rgb}{0.1058824,0.5215686,0.06666667}

\definecolor{dkgreen}{rgb}{0,0.6,0}
\definecolor{gray}{rgb}{0.5,0.5,0.5}
\definecolor{mauve}{rgb}{0.58,0,0.82}

\begin{document}
\title{Programming the Adapteva Epiphany 64-core Network-on-chip Coprocessor}


\author{\IEEEauthorblockN{Anish Varghese, Bob Edwards, Gaurav Mitra, Alistair P Rendell}
\IEEEauthorblockA{Research School of Computer Science\\
Australian National University\\
Canberra, Australia\\
Email: anish.varghese@anu.edu.au, bob@cs.anu.edu.au, \\ gaurav.mitra@anu.edu.au, alistair.rendell@anu.edu.au}
}
%\and
%\IEEEauthorblockN{Alistair Rendell}
%\IEEEauthorblockA{Research School of Computer Science\\
%Australian National University\\
%Canberra, Australia\\
%Alistair.Rendell@anu.edu.au}
%}


%Telephone: (800) 555--1212, Fax: (888) 555--1212}


\maketitle

\begin{abstract}
In the construction of exascale computing systems energy efficiency and power
consumption are two of the major challenges. Low-power high performance
embedded systems are of increasing interest as building blocks for large scale
high-performance systems. However, extracting maximum performance out of such
systems presents many challenges.  Various aspects from the hardware
architecture to the programming models used need to be explored. The Epiphany
architecture integrates low-power RISC cores on a 2D mesh network and promises
up to 70 GFLOPS/Watt of processing efficiency.  However, with just 32 KB of
memory per eCore for storing both data and code, and only low level inter-core
communication support, programming the Epiphany system presents several
challenges. In this paper we evaluate the performance of the Epiphany system
for a variety of basic compute and communication operations.  Guided by this
data we explore strategies for implementing scientific applications on memory
constrained low-powered devices such as the Epiphany. With future systems
expected to house thousands of cores in a single chip, the merits of such
architectures as a path to exascale is compared to other competing systems.  

\end{abstract}


% \fi
\IEEEpeerreviewmaketitle

\section{Introduction}
\label{intro} 

The Epiphany architecture comprises a low power, multi-core, scalable,
parallel, distributed shared memory embedded system created by
Adapteva\cite{Epiphanydocuments}. The Epiphany IV 64-core Network-on-chip (NoC)
coprocessor contains 64 cores (referred to as eCores) organized in a 2D mesh
with future versions expected to house up to 4096 eCores. The Parallella
System-on-module (SoM) board\cite{Epiphanydocuments} combines the Epiphany IV
chip with a host ARM processor housed in a Zynq System-on-chip. An earlier
development prototype of the Parallella uses an FPGA mezzanine ``daughter''
card (FMC) housing the Epiphany IV, attached to a ZedBoard~\cite{zedboard}. In
this paper we report our evaluation of the hardware characteristics and
software environment of the Epiphany system from the perspective of an
application program developer using the ZedBoard and FMC daughter card setup.

To assess the performance of the Epiphany system we implement a stencil based
scientific application kernel and a parallel matrix multiplication kernel.
Stencil kernels apply regular operations on a grid, and are common to a wide
range of high performance computing applications, and have similar
characteristics to other applications such as image processing. They require
good floating point performance but also fast communications.  Parallelization
is usually via domain decomposition with communications primarily between
adjacent domains. Such applications might be expected to map well to the 2D
mesh topology of the Epiphany coprocessor.  

Multiplication of matrices is a fundamental operation which is used in many
scientific applications. Here, we implement and extend the parallel matrix
multiplication algorithm described by Sapir \cite{yanivmatmul} which
involves data communication between neighbouring cores following Cannon's
algorithm \cite{cannon}. The relatively small memory per core presents some
challenges in implementing this, necessitating careful usage of available
memory buffer space for communication between cores. 

In the following sections we give a brief overview of the Epiphany system and
the programming model it supports. We then discuss how this impacts on program
design. In Section \ref{performance_char} we outline micro-benchmarks used to
assess the basic performance of the Epiphany coprocessor. Section
\ref{stencil_imp} details the parallel stencil implementation along with
performance results. Section \ref{matmul_main} details the parallel matrix
multiplication implementation along with performance results. Section
\ref{related_work} highlights related work and contrasts the Epiphany with
similar many-core energy efficient systems. Conclusions and future work are
outlined in Section \ref{conclusion}.

\section{System Architecture}

\begin{figure}[ht]
\centering
\includegraphics[width=\columnwidth]{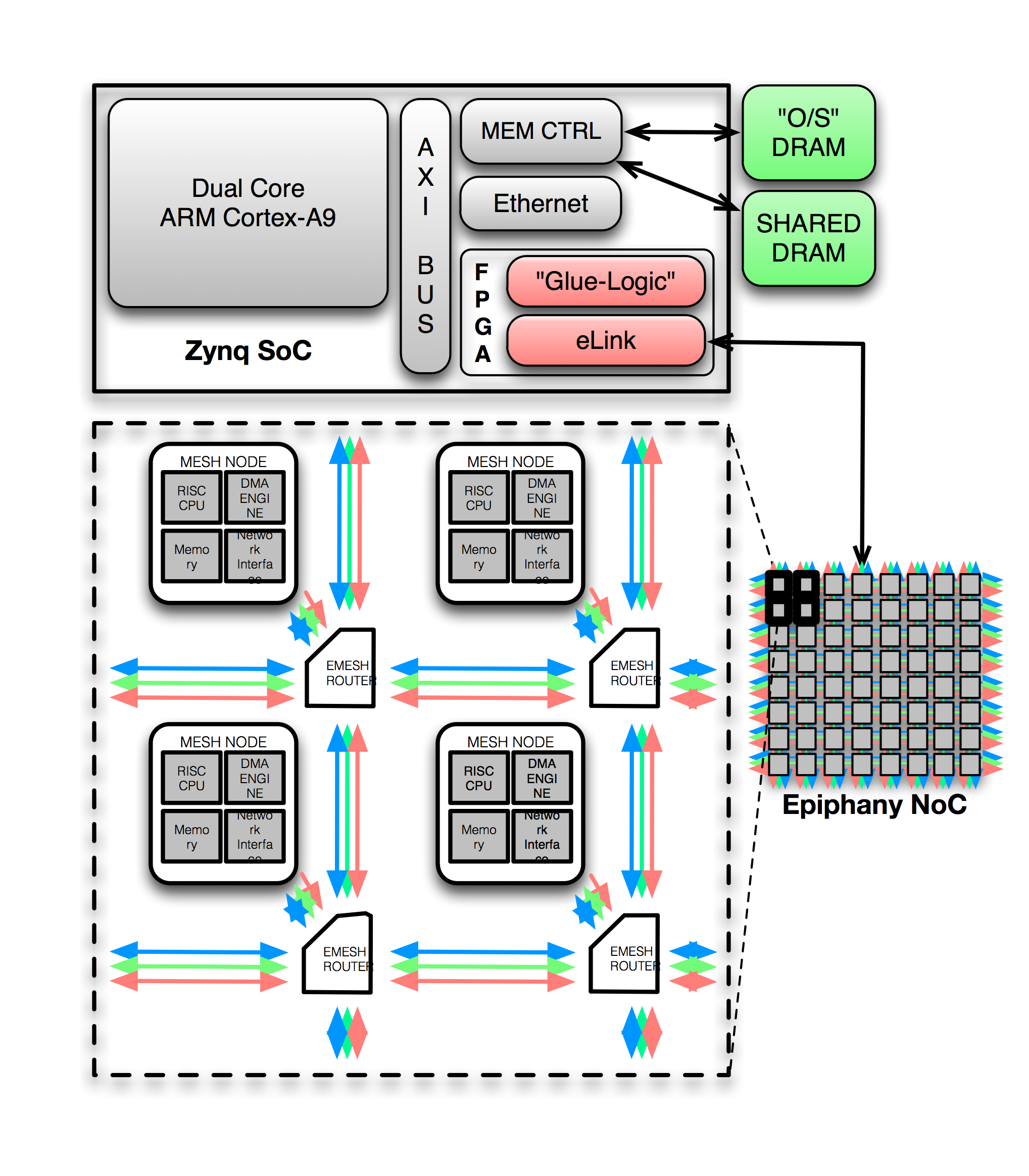}
\caption{Adapteva Epiphany System}
\label{fig:parallella}
\end{figure}

Both the Parallella and the prototype ZedBoard consist of a \emph{Zynq} SoC,
shared memory and the Epiphany NoC coprocessor as shown in
Figure~\ref{fig:parallella}. 

%\subsection{Zynq SoC}

The Xilinx Zynq 7000 series SoC contains a dual-core ARM Cortex-A9 CPU running
at 800 MHz on the Parallella and 667 MHz on the ZedBoard, with standard on-chip
peripherals such as USB 2.0, Ethernet, UART, MIO, AXI BUS, GPIO, HDMI, JTAG
etc. It also contains a Field Programmable Gate Array (FPGA) which is used to
implement the ``Glue-Logic'' and eLink protocol required to interface with the
Epiphany coprocessor. In addition, the FPGA implements the AXI master
interface, AXI slave interface and a HDMI controller.  

%\subsection{Epiphany NoC coprocessor}

The Epiphany NoC has a 2D array of \emph{eCores} connected to each
other by a mesh network-on-chip. Each eCore consists of a RISC CPU, 32 KB of
local scratchpad memory, a Direct Memory Access (DMA) engine,  and a network
interface to an eMesh router. No cache is present. %on the eCores. 
Each eMesh
router provides three network communication channels; % to an eCore as shown in Fig~\ref{fig:parallella}, 
an on-chip write network (in blue), an off-chip
write network (in green) and a read request network (in red). The 
eCore CPU is super-scalar and can execute two floating-point operations and a
64-bit memory load/store operation in every clock cycle. Scratchpad memory % in each eCore 
can theoretically provide up to 32 Bytes per clock cycle of
bandwidth.

%\subsection{Memory Hierarchy}

The Parallella SoM has 1 GB of DDR3 RAM, while the ZedBoard has 512 MB. The
DRAM is partitioned such that Linux running on the ARM Cortex-A9 CPU
has its own private O/S memory and the rest is accessible by both the ARM and
Epiphany.  Shared memory access for the Epiphany is handled by an eLink
interface via the AXI bus and memory controller on the Zynq SoC. The Epiphany
has a flat and unprotected memory map. Each eCore can address its local SRAM,
other eCores' SRAMs and shared off-chip DRAM.

\section{Programming Model}
\label{programming_model}
The Epiphany chip can be programmed using C, and has an
SDK~\cite{Epiphanydocuments} that provides some basic programming primitives to
facilitate writing parallelized C code for this architecture.  Some of the key
features of the SDK are:

\begin{itemize}
    \item \emph{Workgroup model:} To program the eCores,
        workgroups are created by specifying the number of rows and columns of
        nodes and the location of the starting node of the group.  The SDK
        provides functions to determine the ID and location of neighbouring
        eCores.

    \item \emph{Memory addressing:} All eCores share the same address space and
        it is possible to read and write directly to the local memory of
        another eCore. The SDK provides functions to obtain the global address
        of a memory location in another eCore's local memory facilitating data
        transfer between the nodes.

    \item \emph{Communication between eCores:} The SDK provides APIs to
        transfer blocks of data between nodes and to the shared memory. These
        can be achieved by using either the CPU or the DMA engine. Two DMA
        channels are available in each node supporting both non-blocking and
        blocking DMA transfers.

    \item \emph{Barriers:} The SDK provides functions for setting
        synchronization points and barriers in the program.  

    \item \emph{Hardware Mutex:} Mutexes are available to ensure mutual
        exclusion while accessing shared resources. The workgroup defines a
        memory location on the chip as the mutex object. The SDK provides
        functions to enable the eCores to utilize the mutex object.

    \item \emph{Event Timers:} Each eCore has two event timers that can be
        configured to independently monitor key events within the node such as
        clock counter, watchdog timing etc. This can be used to count
        the number of clock cycles which have elapsed 
        during execution of a block of code.
    
\end{itemize}

The steps required to execute a program are: 
\begin {enumerate}
\item Host program creates a workgroup by specifying the number of rows and
    columns required and the position of the start node in the group.

\item Host resets all the nodes and loads the device-side executable
  image into each eCore.

\item Host signals all the eCores in the workgroup to start execution

\item Host communicates with each eCore either by accessing the core's local
    memory or using the shared memory. 

\item Once the execution is complete, the host is signalled. The host reads the
    result either directly from each eCore's local memory or from the shared
    memory.

\end {enumerate}

%% use '\input{Programming.tex}' to include in main document

\section{Programming Considerations}
The Epiphany eCore architecture presents some interesting challenges to
implementing high performance numerical codes. The main limitation is the
relatively small 32KBytes of local RAM per eCore which must be divided between
program code, data and stack. Although each eCore has access to the entire
32-bit address space, performance drops off when accessing non-local memory.
Within each eCore the supplied linker scripts allow the programmer to control
which parts of the code and data are to reside in which specific bank of local
memory and which parts are to be located in slower off-chip shared memory.

In its current form the Epiphany eCore does not include hardware support for
integer multiply, floating point divide or any double-precision floating point
operations. This design decision frees up silicon for other uses, e.g. for
additional cores. Obviously the implication of this varies from application to
application.

Maximum floating-point performance is achieved when each eCore is performing a
stream of Fused-Multiply-Add (FMADD) instructions with simultaneous 64-bit load
or store operations in each clock cycle. At 600MHz on a 64-core Epiphany this
corresponds to a peak of 76.8 single-precision GFLOPS. The ability of the
compiler to optimise code and achieve this is another matter.

\subsection{Program Structure}
%% The Epiphany mesh is intended to operate as a co-processor of a more general-purpose host CPU.
%% In the case of the Parallella, the host CPU is a Xilinx Zynq device with dual ARM A9 CPUs, set up to run Linux.
%% Data moves between the host and the Epiphany through a shared memory.
%% In the case of the Parallella, this is a 32MByte chunk of the 1GByte of external RAM.

The development environment requires (at least) two C programs to be written:
one for the host CPU and one or more ``kernels'' for running on the eCore
nodes. 

%% Library routines facilitate the control of the Ephiphany device, including partitioning the available eCore nodes between one or more control groups.

An application code would typically perform all of its initialization and outer
loops on the host CPU, with the inner-most, numerical-intensive, loops running
as kernels on the Epiphany eCore nodes. Lesser-used library functions required
by the kernel can be located in slower, but more abundant, shared-memory. 

\subsection{Overcoming Memory Limitations}
As indicated previously, the local eCore memory is implemented as four banks of
8KBytes each. Maximum performance can be obtained only when code is fetched
from one bank whilst load/store and DMA operations are occurring to other
banks. 

This further restricts code size to 8 or 16KBytes, or between 2k and 8k
instructions (depending on mix of 16-bit and 32-bit instruction words). The
programmer needs to carefully allocate the use of these four local memory banks
in order to achieve the best performance. 

For example, the programmer could allocate one bank of local memory for code,
two for data (``data1'' and ``data2'') and one for the stack and local
variables. With such an arrangement the code can process data to/from ``data
1'', while using DMA to move data in/out of ``data 2''. When the processing and
DMA are complete, the code can then go on to process ``data 2'' while using DMA
to move result data out and new input data into ``data 1''. 

Adding further pressure on limited memory, branching (eg. in loops) costs 3
cycles, so should be avoided where possible by ``unrolling'' inner loops.
However unrolling loops comes at a cost to code size. With such small amounts
of memory available for code, it is necessary to finely tune the degree to
which loops are unrolled. Directives to the C compiler can be used to determine
the degree of loop unrolling. 

Instructions can, however be fetched from the local memory of other eCores.
Thus a novel approach may be to locate smaller fragments of non-innermost-loop
code in unused portions of banks of local memory of eCores within a row. This
code could then be executed, when required, with contention only between the
eCores in that row. This would result in less contention for eMesh bandwidth
than if all the eCores were executing code out of external shared memory. 

\subsection{Hardware/Software Operation}

Codes for array processing often make use of product terms in array indices,
for example, to calculate row offsets. Without hardware support for integer
multiplication it is desirable to iterate through array elements in a regular
matter using incremented offsets. Similarly where possible floating-point
divide operations should be removed from inner loops or minimized. In both
cases these are optimisations that can usually be carried out by a compiler.

In terms of the current lack of support for double-precision floating point
arithmetic, there is really no sensible work-around. However, there is
increasing evidence to suggest that for many calculations careful use of single
precision floating point is
sufficient~\cite{jd_singleprecision,buttariexploiting}. Also in the case of the
Parallella and other Epiphany platforms, the dual-core ARM Cortex-A9 CPUs on
the Zynq chip provide high-performance double-precision Vector Floating Point
Units (VFPU), so codes with high single-precision requirements and more modest
double-precision requirements may fit the Zynq-Epiphany combination well. 

%% The local memory of the eCore CPUs has equivalent performance to the "level 1" cache of other CPUs, such as the L1 caches of the ARM Cortex-A9 cores on the Zynq host device. However, in this case, the local memory content needs to be manually managed by the programmer, not a cache controller.
%% Furthermore, the local memory is banked into four 8kbyte banks.
%% Accessing code or data from memory locations not local to an eCore requires sending read and/or write requests across the eMesh network to other eCore nodes, or to off-chip shared memory. All such accesses will require contention arbitration.
%% The per-eCore DMA controller can simultaneously transfer data to/from one block of local memory, whilst instructions are fetched from another and load/store operations occur to yet another.

%% A pair of 8-bit channels data paths (an "eLink") is implemented in the Zynq FPGA structure to provide up to 1.4GByte/sec (for Parallella-64) bi-sectional bandwidth to the Epiphany mesh.

\section{Performance Experiments}
\label{performance_char}
%The following experiments were performed in order to evaluate the communication
%capabilities of the Mesh network of the Epiphany (eMesh). 

%\subsection*{Experiment Platform}

{\bf Experimental Platform:} A ZedBoard~\cite{zedboard} evaluation module containing a Xilinx Zynq 7000 SoC
XC7Z020-CLG484-1 with a daughter card~\cite{Epiphanydocuments} housing the Epiphany-IV
64-core 28nm (E64G401) was used for all experiments. The dual-core ARM Cortex-A9
present on the Zynq SoC runs at 667 MHz and the Epiphany eCores run at 600 MHz
each. The board has 512 MB of DDR3 RAM which has 480 MB allocated as ARM Linux
O/S private memory and 32 MB allocated as shared memory between ARM Linux and
Epiphany eCores. 

%\subsection*{Compilers and tools}

{\bf Compilers and Tools:} The Epiphany SDK version 5.13 was used to perform all experiments. The ARM
Cortex-A9 runs the Linux kernel version 3.3.0 and the Linaro root file system
version 12.11. ARM GCC 4.6.3 is used to compile the host code and E-GCC 4.8.2
(Epiphany GCC) is used to compile the device code. Several compiler options are
enabled for E-GCC, including {\tt -funroll-loops -falign-loops=8
    -falign-functions=8 -fmessage-length=0 -ffast -math -ftree-vectorize
    -std=c99 ffp-contract=fast -mlong-calls -mfp-mode=round-near est -MMD -MP
-fgnu89-inline}.
These enable optimizations including and not limited to loop unrolling,
inlining functions etc.

\subsection{Network Performance}
The eMesh Network-On-Chip has a 2D mesh topology with only nearest-neighbour
connections. To evaluate the cost of routing messages from one eCore to another
a small micro-benchmark was written. In this benchmark one eCore in the mesh
writes data as a sequence of 32-bit transfers into the memory of another eCore.
Once the transfers are complete, the source eCore writes to a specific location
in the receiving eCore. The receiving eCore monitors this location, observes the
change, and begins to write the data into the memory of the next eCore in the
row. This process is repeated for all the mesh nodes with the boundary nodes
transferring the message to the next row. This is repeated a number of times
while the total data transferred and total mean time are recorded. Two methods
are used to transfer the data between the two eCores - DMA and point-to-point
writes. Pseudo code for the benchmark with point-to-point write transfers is
given in Listing \ref{bandwidth_code}.

\begin{lstlisting}[language=C,frame=single,basicstyle=\ttfamily\scriptsize,caption=Code
for Message transfer between nodes,label=bandwidth_code]
e_ctimer_set(E_CTIMER_0, E_CTIMER_MAX);
e_ctimer_start(E_CTIMER_0, E_CTIMER_CLK);
time_e = e_ctimer_get(E_CTIMER_0);
for (loopcount=1;loopcount<=LOOP;loopcount++) {
    //Waiting for previous core to finish
    //writing
    while (*flag<loopcount); 

    *val_next0 = *val0 ;       
    *val_next1 = *val1 ;       
    ......
     *val_next19 = *val19 ;     

    *flag_next_core = (coreid!=end_core)?loopcount:loopcount+1;
}
time_s = e_ctimer_get(E_CTIMER_0);
e_ctimer_stop(E_CTIMER_0);
clocks = time_e - time_s;

\end{lstlisting}

\begin{figure}[ht]
\centering
\includegraphics[width=3.7in]{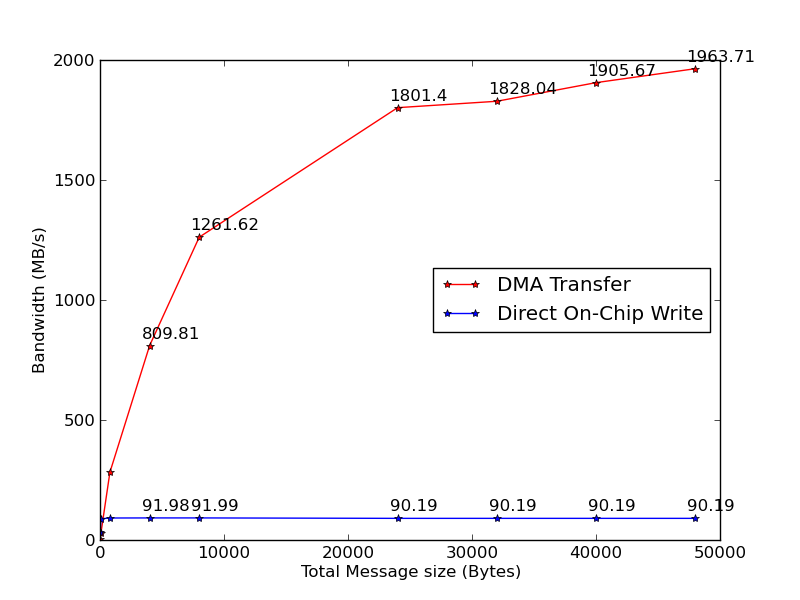}
\caption{Bandwidth - DMA vs Direct Writes}
\label{bandwidth_graph}
\end{figure}

The bandwidths observed using the DMA and direct write methods as a function of
message length for transfers between adjacent eCores are shown in Figure
\ref{bandwidth_graph}. For all but very small messages it is better to use DMA
rather than issuing individual write instructions. For large messages DMA is
able to achieve around 2GB/s transfer rates. Theoretically, with a 32-bit
single word transfer per clock cycle, the DMA engine can provide a sustained
data transfer rate of 2.4GB/sec at a clock speed of 600 MHz. With doubleword
transfers it can provide a transfer rate of up to 4.8GB/sec.

\begin{figure}[ht]
\centering
\includegraphics[width=3.7in]{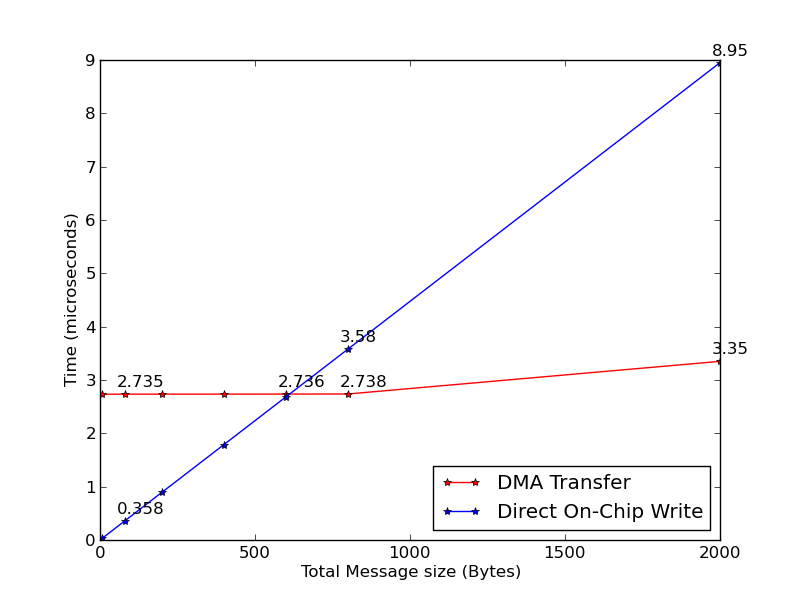}
\caption{Latency - DMA vs Direct Writes}
\label{latency_graph}
\end{figure}

Latency is important for small data transfers. Figure \ref{latency_graph} shows
the latency for small message transfers. For transfers of less than about 500
bytes it is faster to write directly into the memory of an adjacent eCore
rather than to use DMA transfers.  Beyond 500 bytes, DMA is preferable.

In Table \ref{routing_table} we report the latency for an 80 byte message
transferred from eCore 0,0 to one of the other cores in the $8 \times 8$ grid.
The Manhattan distance of each transfer is given. This shows surprisingly
little effect of distance, although all transfers are relatively slow in terms
of clock cycles.

\begin{table}[ht]
\begin{center}
\begin{tabular}{|c|c|c|c|}
\hline 
Node 1 & Node 2 & Manhattan Distance & Time per transfer (nsec) \\
\hline
0,0 & 0,1  & 1 & 11.12 \\
0,0 & 1,0  & 1 & 11.12 \\
0,0 & 0,2  & 2 & 11.14 \\
0,0 & 1,1  & 2 & 11.14 \\
0,0 & 1,2  & 3 & 11.19 \\
0,0 & 3,0  & 3 & 11.19 \\
0,0 & 0,4  & 4 & 11.38 \\
0,0 & 1,3  & 4 & 11.38 \\
0,0 & 3,3  & 5 & 11.62 \\
0,0 & 4,4  & 6 & 11.86 \\
0,0 & 7,7  & 14 & 12.57 \\
\hline 
\end{tabular} 
\caption{Effect of Node distance on Transfer Latency}
\label{routing_table}
\end{center}
\end{table}

\subsection{External Shared Memory}
\label{ext_mem_perf}
As mentioned, the only way to get data in and out of the Epiphany chip is via
the shared memory (unless external hardware is connected to the other eLink
interfaces).

Much example code exists to showcase the performance of the memory system, but
none to clearly show the performance when multiple eCores attempt to write to
the external shared memory, over the single 8-bit wide, 600MHz (600MB/sec each
direction) eLink, and how these accesses may be impacted by normal ARM CPU
memory accesses to the shared DRAM.

Testing on the 64-core Epiphany-IV device housed on our system was hampered by
the presence of Errata \#0, as documented in the E64G401
Datasheet~\cite{Epiphanydocuments}: ``Duplicate IO Transaction''.  This is
reported to affect all eCores in row 2 and column 2 (15 eCores total) in the
64-core device, for instruction fetches and data reads, but not for DMA
operations, nor, apparently, for data writes.

Testing showed that location does matter when an eCore is attempting to write
to the external shared memory. Nodes closer to column 7 and row 0 get the best
write access to external DRAM. Nodes closer to column 7 always do better than
eCores closer to row 0. With sufficient contention, many (all) eCores in rows 5 -
7 simply miss out on write slots.

In any case, the maximum write throughput to external shared memory achieved
was 150MB/sec, exactly one quarter of the theoretical maximum of the 600MB/sec
eLink.

To evaluate the relative share of the external memory interface that is
allocated to each eCore for off-chip data transfers, a micro-benchmark was
written. In this benchmark, four eCores (organized as $2 \times 2$)
continuously write blocks of 2KBytes as sequences of 4-byte stores to the
external memory.  This is done for a specific period of time (two seconds) and
the utilization of the eLink by each mesh node is measured.  The result is
shown in Table \ref{ext_mem_perf_table1}.  The experiment is repeated with all
64 eCores writing simultaneously to the external memory and the results are
shown in Table \ref{ext_mem_perf_table2}. The effects of starvation are clearly
evident. 

\begin{table}[ht]
\begin{center}
\begin{tabular}{|c|c|c|}
\hline 
Mesh Node & Iterations & Utilization  \\
\hline
0,0 & 61037 & 0.41 \\
0,1 & 48829 & 0.33 \\
1,0 & 24414 & 0.17 \\
1,1 & 12207 & 0.08 \\
\hline
\end{tabular} 
\caption{4 Mesh Nodes writing 2KB blocks to DRAM over 2 seconds}
\label{ext_mem_perf_table1}
\end{center}
\end{table}

\begin{table}[ht]
\begin{center}
\begin{tabular}{|c|c|c|}
\hline 
Mesh Node (Total No) & Iterations & Utilization  \\
\hline
0,7 1,7 2,7 3,7 & 27460+ & 0.187 each \\
(8) & 3050+ & 0.021 each \\
(4) & 2040+ & 0.014 each \\
(8) & 100 - 1000 &  \\
(9) & 10 - 100 &  \\
(7) & 1 - 10 &  \\
(24) & 0 & \\
\hline 
\end{tabular} 
\caption{64 Mesh Nodes writing 2KB blocks to DRAM over 2 seconds}
\label{ext_mem_perf_table2}
\end{center}
\end{table}

\section{Heat Stencil}
\label{stencil_imp}

We use the same stencil as in the Intel Teraflop paper~\cite{mattsonintel}. In
this benchmark a 5-point star (``+'') shaped stencil, with separate
co-efficients per point, is applied to all data points in the grid.  We
reference the five points as Top, Left, Centre, Right and Bottom (T,L,C,R,B).
Using $i$ and $j$ to reference points on a 2D Cartesian grid in the $x$ and $y$
directions and $T$ as the temperature, an update proceeds as follows:

\begin{dmath*}
    \label{eq:heat}
    T_{new_{i,j}} = w_1 * T_{prev_{i,j+1}} + w_2 * T_{prev_{i,j}} \\
    + w_3 * T_{prev_{i,j-1}} + w_4 * T_{prev_{i+1,j}} \\+ w_5 * T_{prev_{i-1,j}}
\end{dmath*}

The stencil kernel is mapped to the Epiphany architecture using a 2-dimensional
domain decomposition. The grid of temperatures is stored in a 1-dimensional
array in row-major order and is distributed equally among all the nodes. The
host transfers the corresponding grid portion to the local memory of each eCore
directly using the available API functions for data transfer. Once the grid is
copied to the local memory, each eCore computes the values for the current
iteration for all the grid points assigned to that eCore. This is followed by a
communication phase.

\subsection*{Computation}
\label{computation_phase}
Maximum floating point performance on the Epiphany architecture can only be
achieved when using the FMADD instructions which effectively yields two
Flops/cycle.  This instruction multiplies two inputs from registers and
accumulates the result into a third register, all in one instruction. It can be
executed concurrently with certain other integer unit instructions, such as
loads and stores, in a super-scalar manner. 

\subsection*{Communication}
\begin{figure}[ht]
\centering
\includegraphics[width=3.5in,keepaspectratio]{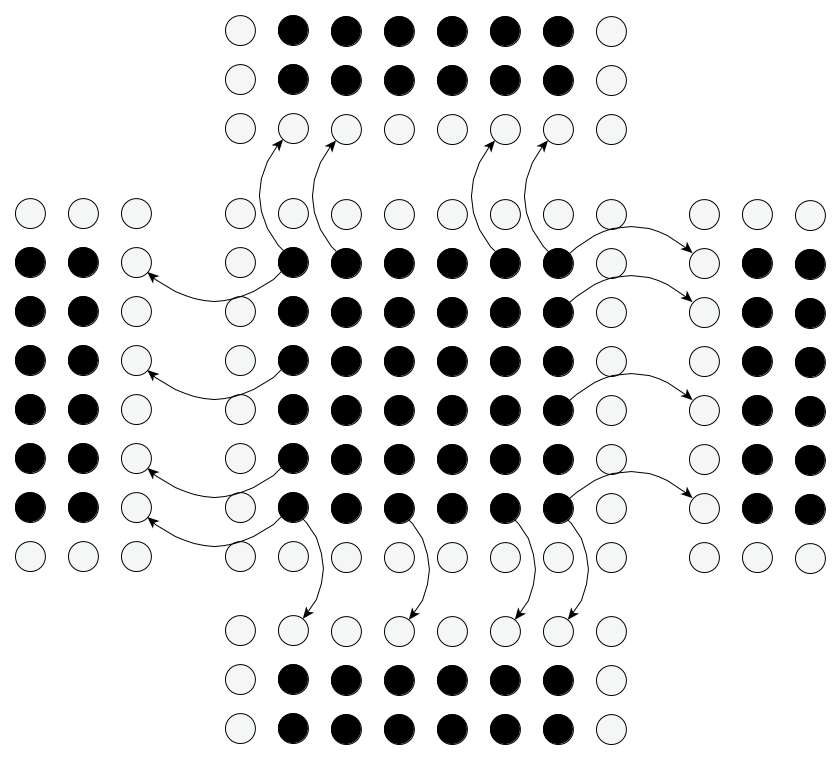}
\caption{Communication of boundary data}
\label{stencil_communication}
\end{figure}

The computation is followed by a communication phase where the ``edge'' regions
of the grid are transferred to the ``boundary'' regions of each of the four
neighbouring eCores. Thus each eCore receives data from each of its neighbours
as shown in Figure \ref{stencil_communication}. An ``in-place'' algorithm is
used where the result of the current iteration is stored back in the same
array.  Hence the transfers are started after the neighbours have completed
their computation phase. In each iteration, a node is synchronized with each of
its four neighbouring nodes. The transfers are achieved using the DMA engine,
which transfers data to each neighbour in a chain. 64-bit double word transfers
are used for the top and bottom boundary rows as they are stored in continuous
memory locations, while 32-bit single word transfers are used for transferring
the left and right boundary columns.

The following snippets of code illustrates how the communication and synchronization
are performed.

\begin{lstlisting}[language=C,frame=single,basicstyle=\ttfamily\scriptsize,caption=Code
for Communication and Synchronization]
//Defining DMA Descriptors
start_descr0=start_descr1=0x0000;
dma_config = E_DMA_ENABLE | E_DMA_MASTER;
config_row = dma_config | E_DMA_DWORD;
config_col = dma_config | E_DWA_WORD;
//BOTTOM
if (core_row != group_rows - 1) {
    dst_offset = 0;
    src_offset = (CORE_GRID_Y - 2) * CORE_GRID_X;
    e_dma_set_desc(E_DMA_0, config_row, start_descr0,
       0x0008, 0x0008,
       CORE_GRID_X>>1, 0x0001,
       0x0000 , 0x0000,
       (void *)(dptr+src_offset),(void *)(t_neighbour[BOTTOM]+dst_offset),&dma_desc[3]);
    start_descr0=&dma_desc[3];
}
//TOP
if (core_row != 0) {
    dst_offset = (CORE_GRID_Y - 1) * CORE_GRID_X;
    src_offset = CORE_GRID_X;
    if (start_descr0!=0x0000) config_row|=E_DMA_CHAIN;
    e_dma_set_desc(E_DMA_0, config_row, start_descr0,
       0x0008, 0x0008,
       CORE_GRID_X>>1, 0x0001,
       0x0000 , 0x0000,
       (void *)(dptr+src_offset),(void *)(t_neighbour[TOP]+dst_offset),&dma_desc[2]);
    start_descr0=&dma_desc[2];
}
//RIGHT
if (core_col != (group_cols - 1)) {
    dst_offset = 0;
    src_offset = CORE_GRID_X - 2;
    e_dma_set_desc(E_DMA_1, config_col, start_descr1,
       0x0000, 0x0000,
       0x0001, CORE_GRID_X, 
       (CORE_GRID_X*sizeof(float)),(CORE_GRID_X*sizeof(float)),
       (void *)(dptr+src_offset),(void *)(t_neighbour[RIGHT]+dst_offset),&dma_desc[1]);
    start_descr1=&dma_desc[1];
}
//LEFT
if (core_col != 0) {
    dst_offset = CORE_GRID_X - 1;
    src_offset = 1;
    if (start_descr1!=0x0000) config_col|=E_DMA_CHAIN;
    e_dma_set_desc(E_DMA_1, config_col, start_descr1,
       0x0000, 0x0000,
       0x0001, CORE_GRID_X,
       (CORE_GRID_X*sizeof(float)),(CORE_GRID_X*sizeof(float)),
       (void *)(dptr+src_offset),(void *)(t_neighbour[LEFT]+dst_offset),&dma_desc[0]);
    start_descr1=&dma_desc[0];
}

//Synchronize and Transfer data per iteration
iter++;                                                        
*(iter_neigh[LEFT]) = iter;
*(iter_neigh[RIGHT]) = iter;
*(iter_neigh[TOP]) = iter;
*(iter_neigh[BOTTOM]) = iter;            
while (iter_array[TOP]<iter||iter_array[BOTTOM]<iter   
  ||iter_array[LEFT]<iter||iter_array[RIGHT]<iter);  

//Start dma
e_dma_start(start_descr0,E_DMA_0);
e_dma_start(start_descr1,E_DMA_1);
e_dma_wait(E_DMA_0);
e_dma_wait(E_DMA_1);
//End of dma

*(t_iter_neigh[LEFT]) = iter;
*(t_iter_neigh[RIGHT]) = iter;
*(t_iter_neigh[TOP]) = iter;
*(t_iter_neigh[BOTTOM]) = iter;             
while (t_iter_array[TOP]<iter||t_iter_array[BOTTOM]<iter    
  ||t_iter_array[LEFT]<iter||t_iter_array[RIGHT]<iter);
\end{lstlisting}

The above steps are repeated for a number of iterations. After all the
iterations are completed, the host reads the corresponding portion of the
computed grid from each eCore and writes out the final result.

\subsection*{Discussion}
Our first implementation was written in C, however the relatively immature
compiler was only able to achieve a small fraction of peak. This code was
replaced with a hand-tuned assembly code using the Epiphany instruction set.
Grid sizes of $20 \times X$ were used where 20 was chosen based on the number
of available registers and the latency of operations. Rows containing more than
20 elements are processed 20 at a time. The maximum number of rows, X, that can
be processed on one eCore is driven by the number of elements per row and the
available memory.

Experimentation showed that the register used for accumulating the result of
the FMADD instruction cannot be used again as a Floating point unit (FPU)
source or result register, or as the source of a store instruction for at least
5 cycles to avoid stalling the execution pipeline.  

The eCore CPU has a total of 64 accessible 32-bit registers which can be used
as single-precision floating point values, 32-bit signed or unsigned integers
or as memory pointers, with various addressing modes.  The Epiphany
Architecture Reference manual recommends that some of these registers be used
for parameter passing. Some registers are to be saved by the calling routine
and some are to be saved by the called (``Callee'') routine.  The only special
register is r14, the Link Register, which contains the return address used in a
``ReTurn'' from Subroutine (RTS) instruction.

Experimentation with the e-gcc C compiler showed that the registers identified
as requiring ``Callee Saved'' (22 of them) are only available if the special
word ``register'' is prepended to local variable declarations. The four
registers identified as ``Reserved for constants'' are not allocated and hence
not used in any C code fragments we inspected.

\subsection*{Attaining Peak Performance}
To attain peak FPU performance for the 5-point stencil, it is desirable to
execute FMADD instructions for as long as possible.  Branching costs 3 cycles,
with a further cycle or two for decrementing a counter register.  Therefore,
inner loops should be unrolled as much as possible, modulo code memory size
constraints.

We maximize the use of registers by buffering rows of input data into registers
and accumulating the results in registers before writing out the final result.
Our strategy is to buffer two rows of grid points whilst performing the five
FMADD instructions per grid point.  We use row lengths (stripes) of 20 points
(a multiple of 5) and enforce a design goal that each grid data point is loaded
into a register just once.

Five FMADD operations are performed on five consecutive T grid points, followed
by five FMADDs on the respective L values, which is followed by the five C
values, five R values and finally the five B values. After completing a run of
five grid points, the accumulated results need to be saved and the accumulators
cleared.  This takes 10 cycles.

To avoid stalling the FPU, we immediately start a second run of five grid points,
using a second set of five accumulators. This is effectively double-buffering the
result accumulators, using 10 registers in total (r8 - r12 for the first set,
r15 - r19 for the other).

During the execution of these 5 x 5 FMADD instructions, we use the ``spare''
integer operation slots to replace the five Top (T) grid points with the five
Bottom (B) grid points, whilst leaving the seven middle (five each of Left,
Centre, Right, with overlap) buffered values alone.  We also use these spare
slots to save the accumulated results from the previous five grid points and to
clear the next five accumulators.

\subsection*{Use of row stripes}
As mentioned earlier, we use row stripes of 20 points. Two rows of 20 data
points, plus the ``boundary'' values at each end, requires a total of 44
registers to buffer the input data.

Before starting, registers r20 - r41 are pre-loaded with grid data for the top
boundary row (T), and registers r42 - r63 with grid data for the middle row
(L,C,R).  As the FMADDs are performed on the five lots of T data buffered in
the registers, and during the FMADDs of the L, C and R grid points, the T data
in r20 - r41 is progressively replaced with the equivalent B data from the next
row of grid data.  These loads need to be complete before the five final FMADDs
on the B data are performed.

At the commencement of the next row, r20 - r41 now contain the middle data (L,
C, R) and r42 - r63 contain the new T data.  During the processing of the
FMADDs for this row, r42 - r63 are progressively replaced with B data from the
next row.  At the completion of the second row, the above registers will be in
the same order as at the start, that is T data in r20 - r41 and L, C, R data in
r42 - r63.

This constitutes one ``unrolled'' loop of 40 x 5 = 200 FMADD instructions and
ideally the same number of cycles.  The code for the loop is approximately
1300 bytes: 800 bytes for the 200 x 32-bit FMADD instructions, plus 480 bytes
for 120 x 32-bit integer instructions performing loads, stores and clears, plus
sundry others.  There is also a 4 or 5 cycle loop ``penalty'' as a register is
decremented and a conditional branch is made to the top of the loop.

\subsection*{Assembly code structure}
Many attempts were made to implement the above operations in C.  However, a
number of issues were encountered.

The main issue was that the C compiler was reluctant to allow all 64 registers
(63 not including the Stack Pointer) to be used. Hence there were a number of
data move instructions in the resulting assembly code to block the dual-issuing
of FPU and integer/data movement instructions.

The main problem with writing the code in assembly language was allocation of
the registers.  Minor code changes could result in large rewrites of register
usage, which inevitably makes the code prone to errors.

To avoid writing too much code, two macros were written to perform each of the
5 x 5 FMADD runs. Calling them alternately, whilst keeping the sequencing of
register numbers correct greatly simplified the code.

Each macro results in 25 FMADD instructions with 15 data movement instructions
interleaved, for a total of 40 x 32-bit instructions, executing in 25 clock
cycles and performing 50 Flops.

Stringing 4 pairs of these macros together results in 200 FMADDs, almost 1300
bytes of code and 400 Flops for two stripes of grid data.

The decrement and branching at the end of a run of two rows of the stripe adds
4 or 5 clock cycles and so a 2 or 2.5\% overhead over 200 clocks.

\subsection*{Other Strategies}
A variation on the design is to allow multiple (3) loads of each grid data
point and to only buffer one row of the "stripe".  This would allow the stripe
to become up to 32 grid points wide, or more.  64-bit loads and stores can be
used to make this possible, but more careful attention to data alignment in
memory and register allocation would be required.

The initial implementation performs the stencil operation ``in place'',
overwriting the previous grid data with new data on each iteration.

By allowing full dual-buffering of the grid data, an assembly language stencil
code with no per-row buffering can be implemented, allowing arbitrary width
only limited by code memory size and per-core data memory size.  The downside
is that this approach requires twice as much memory per grid point, or the grid
array is limited to half the size.

\subsection*{Further Observations}
The assembly code 5-point star stencil can be trivially modified to perform any
5-point stencil within a $3 \times 3$ area containing a grid point, such as a
``X'' shaped stencil, or an arrow.

To change the number of points in the stencil will require some re-writing.
Decreasing the number of points will be relatively straight-forward.  However,
increasing the stencil to, say, a full 9-point stencil will possibly require
some more registers, only currently possible by shortening the stripe width.
It may be possible to load the various stencil co-efficients into registers
from memory, as required, to increase the number of ``points'' per stencil.

\subsection*{Further Optimizations}
At the completion of each iteration, the boundary row/columns of adjoining mesh
nodes need to be updated with ``edge'' data from the new grid, whilst ``edge''
data from surrounding nodes needs to be copied to the boundary row/columns of
the local grid. To do this more efficiently for the ``in-place'' algorithm, the
boundary rows and columns can be double-buffered. This would allow the
transferring of boundary data to neighbouring mesh nodes to commence, whilst
those nodes may still be processing the current boundary data. Performance
gains are likely to be modest, roughly the same as the difference between the
results with and without communication discussed below.

\subsection{Stencil Results} 
\label{stencil_results}
%A number of experiments are conducted to compare the performance of the kernel
%on different configurations of eCores and problem sizes. We also evaluate how
%well the kernel scales to larger problem sizes and increasing number of eCores. 

\subsubsection{Floating Point Performance}

Here, we compare the floating point performance of the stencil kernel for
different configurations of grid sizes. The stencil is evaluated for 50
iterations. Using a row width of 20, as explained in Section
\ref{computation_phase}, we run multiple stripes of $20 \times X$, where X is
the number of rows, one after the other to test larger grid sizes. Three
scenarios are considered i) the performance on a single eCore as a function of
grid size ii) the performance using all 64 eCores when running the same problem
on each eCore, iii) the performance when running one grid across all 64 eCores
including communication of the boundary region between eCores.  

\begin{figure}[ht]
\centering
\includegraphics[width=3.7in]{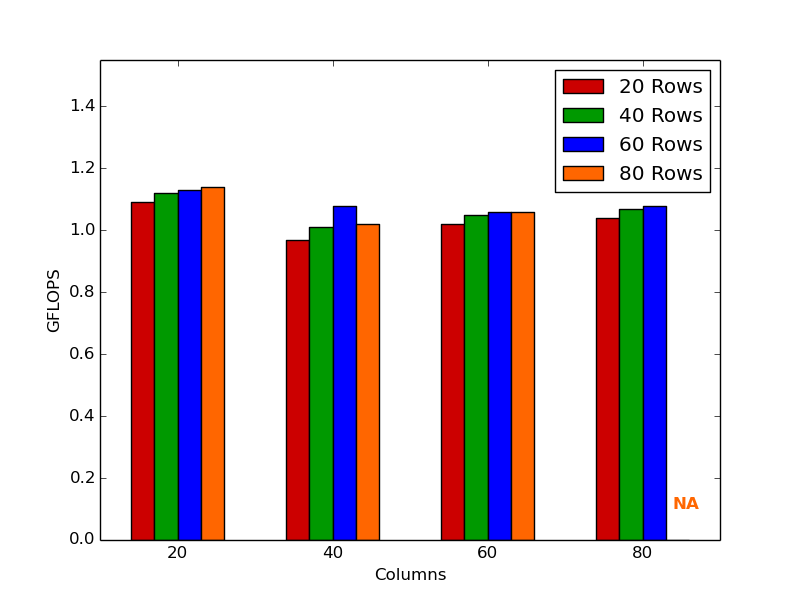}
\caption{Single core Floating point performance}
\label{stencil_singlecore}
\end{figure}

On a single eCore the performance ranges from 0.97-1.14 GFLOPS or between
81-95\% of peak as shown in Figure \ref{stencil_singlecore}. For small sizes,
grids with more rows than columns tend to perform slightly better than the same
size grid with more columns than rows. This is due to the overhead involved in
performing multiple stripes of computation when column size is greater than 20
elements. 

\begin{figure}[ht]
\centering
\includegraphics[width=3.7in]{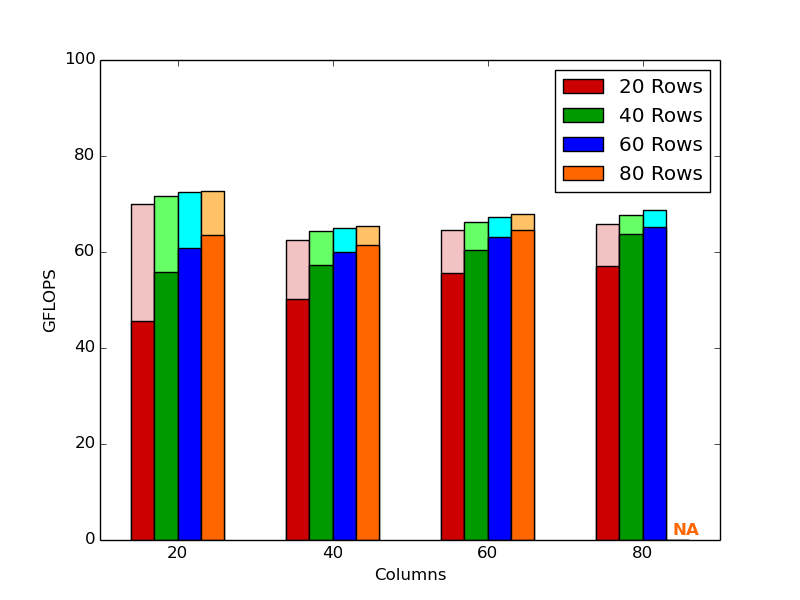}
\caption{64-core Floating point performance}
\label{stencil_multicore}
\end{figure}

The performance of the code on all 64-eCores is shown in Figure
\ref{stencil_multicore}. The darker colours show the performance of the stencil
kernel including communication of boundary region. The lighter colours at the
top of each bar shows the performance without communication of data.

%\begin{table}[ht]
%\begin{center}
%    \begin{tabular}{|p{1.7cm}|c|c|c|c|c|}
%\hline 
%& \multirow{2}{*}{Rows} & \multicolumn{4}{|c|}{Columns} \\\cline{3-6}
%    && 20   & 40   & 60   & 80  \\   
%\hline
%    &  20 & 1.09 & 0.97 & 1.02 & 1.04 \\
%Single core 
%    &  40 & 1.12 & 1.01 & 1.05 & 1.07 \\
%stencil
%    &  60 & 1.13 & 1.08 & 1.06 & 1.08 \\
%    &  80 & 1.14 & 1.02 & 1.06 &  -   \\
%\hline
%64-core stencil
%    &  20 & 69.95 & 62.59 & 64.65 & 65.81 \\
%without
%    &  40 & 71.71 & 64.48 & 66.32 & 67.77 \\
%communication
%    &  60 & 72.45 & 65.14 & 67.21 & 68.85 \\
%    &  80 & 72.83 & 65.47 & 67.99 &  -    \\
%\hline
%64-core stencil
%    &  20 & 45.71 & 50.39 & 55.65 & 57.02 \\
%with
%    &  40 & 55.90 & 57.33 & 60.52 & 63.76 \\
%communication
%    &  60 & 60.86 & 60.09 & 63.19 & 65.28 \\
%    &  80 & 63.60 & 61.58 & 64.51 &   -   \\
%\hline 
%\end{tabular} 
%\caption{Floating point performance in GFLOPS - Stencil computation}
%\label{stencil_flops}
%\end{center}
%\end{table}

As expected, when the computations are replicated across all 64 cores with no
communications performance scales linearly with a peak performance of 72.83
GFLOPS for a stencil containing 80 rows and 20 columns as shown in Figure
\ref{stencil_multicore}. When boundary data is transferred during each
iteration, this performance drops to 63.6 GFLOPS or 82.8\% of peak. Thus the
penalty associated with not overlapping communication and computation is
roughly 9 GFLOPS. Due to the nature of 2D DMA block transfers, grids with more
columns than rows show less performance drop than equivalent grids with more
rows than columns.

\subsubsection{Weak Scaling}
\begin{figure}[ht]
\centering
\includegraphics[width=3.2in]{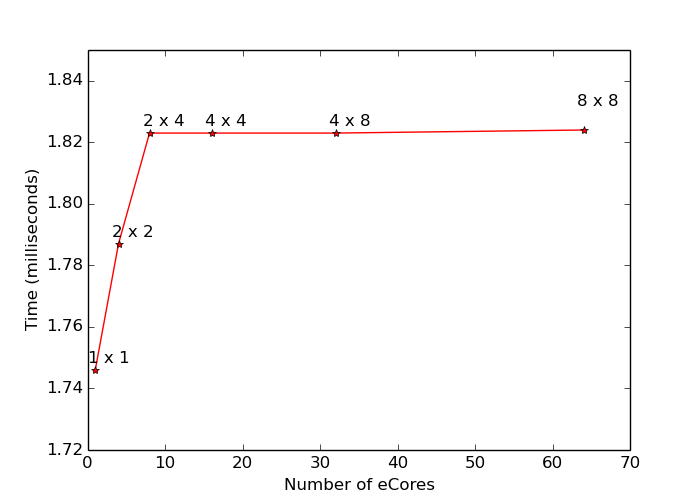}
\caption{Weak Scaling - Number of eCores vs Time}
\label{weak_scaling_graph}
\end{figure}

In this experiment, we increase the number of eCores from 1 to 64. The problem
size is also increased accordingly from $60 \times 60$ (1 eCore) to  $480
\times 480$ (64 eCores). The running time for each configuration is plotted in
Figure \ref{weak_scaling_graph} along with the configuration of the eCores (as
$rows \times columns$). Initially as the number of eCores increases from 1, the
time taken increases due to the need for data communication between the eCores.
This increase quickly levels out after 8 eCores ($2 \times 4$) as communication
between independent pairs of eCores can then be overlapped.

\subsubsection {Strong Scaling}
\begin{figure}[ht]
\centering
\includegraphics[width=3.2in]{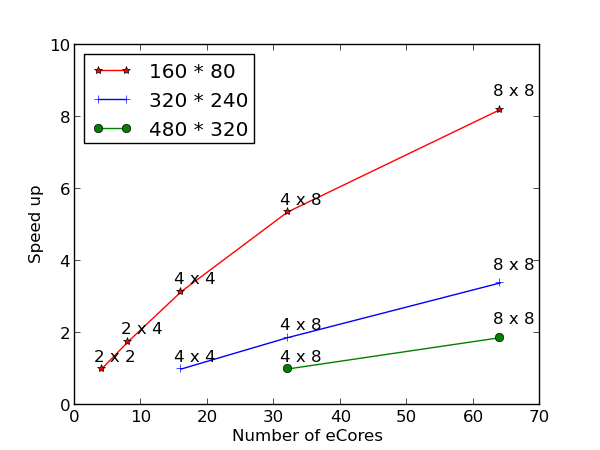}
\caption{Strong Scaling - Number of eCores vs Speedup}
\label{strong_scaling_graph}
\end{figure}

In this experiment, we increase the number of eCores from 1 to 64 while keeping
the problem size fixed. This is repeated for three problem sizes. The speed-up
achieved is shown in Figure \ref{strong_scaling_graph} along with the
configuration of the eCores. In all cases when the number of eCores is first
doubled, a speed up of close to 2 is achieved with slightly better results
achieved for the larger problem sizes. Further doubling of the number of eCores
(when possible) achieves slightly less performance gain.

\section{Matrix Multiplication}
\label{matmul_main}

There are several parallel algorithms for matrix multiplication (matmul) on
many-core systems. The approach used here is based on Sapir~
\cite{yanivmatmul}. Our implementation operates at three levels:

\begin{itemize}
    \item
        At the most basic level, matrix blocks that fit inside a single eCore's
        SRAM are multiplied. The requirement here is for a matrix multiply
        routine that is optimized for a single eCore both in terms of
        performance and memory usage.

    \item 
        At the next level, if the matrices are too large to be stored on a
        single eCore they are block distributed across the memory of multiple
        eCores. The algorithm proceeds by executing the kernel matrix multiply
        on each eCore for a given set of component blocks and then shuffling
        the blocks between eCores, repeating this process until the overall
        matrix multiplication is complete. 

    \item
        At the top level, if the matrices are too large to be stored on the
        entire chip a procedure analogous to the block wise algorithm outlined
        in the previous step is used to orchestrate movement of portions of the
        different matrices between off-chip shared memory and distributed eCore
        memory.

\end{itemize}

Below we expand on each of the three levels. For the purpose of what follows we
consider matrices $A$ and $B$ with dimensions ($M \times N$) and ($N \times K$)
respectively that are multiplied together to form $C$ with dimensions ($M
\times K$).

\subsection*{Tuned single-core matmul kernel}
\label{matmul_singlecore}
As with the stencil code initial attempts were made to develop the eCore matrix
multiply kernel in C. This however gave only 60 \% of peak performance.
Therefore the inner most loop of the matrix multiply was replaced with
hand-tuned assembly code.

The assembly code loads 4 elements of the first matrix (matrix A) into 4
registers (r11,r12,r14 and r15) at a time. In turn each of these elements is
multiplied with each element in the corresponding row of the second matrix
(matrix B) with the intermediate results accumulated into 32 registers
(r32-r63). In this process the rows of matrix B are loaded 8 elements at a time
into registers r16-r23. Double-word loads are used allowing these 8 elements to be 
loaded in 4 clock cycles. 

By pre-loading a few elements of matrix A and B, after each has been used the
next unprocessed element is loaded into the freed registers. This enables load
instructions and FMADD instructions to be interleaved, although care must be
taken to ensure there are at least 5 cycles between using the same register for
a load and a floating point instruction in order to avoid stalling the
execution pipeline.

Each row of matrix A is loaded only once. Each element in the row is multiplied
with all the elements in the corresponding row of matrix B. For example, the
first element in a row of matrix A is multiplied with all the elements in the
first row of matrix B and the intermediate results are accumulated in the first
row of matrix C. The second element in a row of matrix A is multiplied with all
the elements in the second row of matrix B, with the intermediate results being
accumulated in the second row of matrix C. This means that for each row of
matrix A, all the rows of matrix B need to be loaded from memory. Once all the
elements in a row of matrix A are processed, the corresponding row of matrix C
will have its final result. These values are now written out from the
intermediate registers to memory using double-word store instructions and the
registers are cleared for the next row of results.

\subsubsection*{Assembly code structure}
As with the stencil, a macro was written to simplify the code. The macro is
used to multiply an element of matrix A with all the elements in a row of
matrix B. This involves 32 FMADD instructions and around 18 data movement
instructions interleaved, resulting in 50 instructions executing 64 Flops in 32
cycles. For a $32 \times 32$ matmul, the macro is expanded 32 times for
computing each row of product matrix C, resulting in around 6.5 KBytes of
assembly code and 2048 Flops for each row of result. At the end of a row, the
code loops around to compute another row of the result incurring some overhead
for the branch operation. 

The disadvantage of writing in assembly is that the code is not very flexible
to changes to the sizes of the operand matrices, the 'M' dimension of matrix A
being the only parameter which is configurable in the current code (as it is
the loop count). To operate on different sizes of operand matrices, a few
changes would need to be done to the assembly code including the macros in
order to perform matrix multiplication efficiently for those sizes.

\subsubsection*{Memory Considerations}
The operand matrices A and B, and the product matrix C are stored in the local
memory of each eCore. Each eCore stores matrices of sizes up to $32 \times 32$
using a total of 12 KBytes for storing the three matrices. The matrices are
placed in different data banks.  The operand matrices A and B are stored in
data bank 2 and the product matrix C is stored in the last data bank (bank 3).
The entire code takes around 11 KBytes of storage and occupies the
first data bank (bank 0) and portions of the second data bank (bank 1) with the
stack being allocated in the bottom half of bank 1. The size of the code has to
be kept in mind while allocating memory for the operand matrices. This is
especially important for the multi-core matmul version as described below.

\subsection*{On-chip multi-core matmul kernel}
\label{matmul_multicore}
Using the single-core version as a building block, we implement a multi-core
version in order to operate on bigger matrices. With each eCore able to store
operands of sizes $32 \times 32$, we can work on matrices of size $256 \times
256$ with all the data residing in the local memory of the 64 eCores. 

Using capitals to refer to blocks of each matrix, expanding the matrix
multiplication we obtain:    

\begin{dmath*}
    C_{11} = A_{11} B_{11} + A_{12} B_{21} + A_{13} B_{31} + ...
\end{dmath*}
\begin{dmath*}
    C_{12} = A_{11} B_{12} + A_{12} B_{22} + A_{13} B_{32} + ...
\end{dmath*}
\begin{dmath*}
    \vdots
\end{dmath*}
\begin{dmath*}
    C_{21} = A_{21} B_{11} + A_{22} B_{21} + A_{23} B_{31} + ...
\end{dmath*}
\begin{dmath*}
    C_{22} = A_{21} B_{12} + A_{22} B_{22} + A_{23} B_{32} + ...
\end{dmath*}
\begin{dmath}
    \label{eq:matrix_alignment}
    \vdots
\end{dmath}

\begin{figure}[ht]
\centering
\includegraphics[width=4.0in,keepaspectratio]{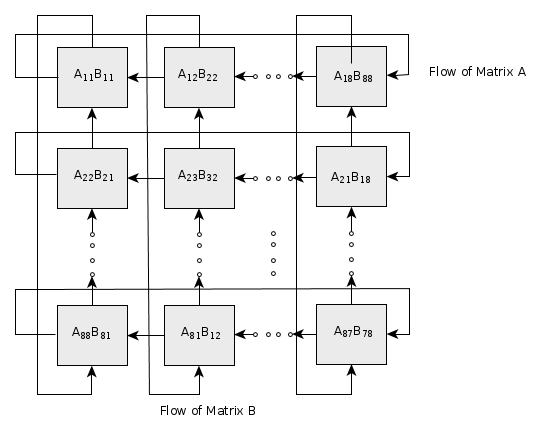}
\caption{Assignment of blocks of A and B and data flow between eCores}
\label{matrix_data_cycling}
\end{figure}

If each eCore is assigned a specific block of $C$, we can see from
equation \ref{eq:matrix_alignment} the blocks that are required by each eCore
in order to complete the matrix product. In the implementation used
here for each matrix a row of blocks is mapped to a row of eCores. The
multiplication proceeds using Cannon's algorithm, where blocks of A
are progressively rotated around rows of eCores while blocks of B are
rotated around columns of eCores. This process is illustrated in
Figure \ref{matrix_data_cycling}. 

For block sizes less than $32 \times 32$, double buffering is used for each of
the operand matrices A and B in order to overlap computation and communication,
thereby improving performance. Once an eCore completes its block computation, it
transfers its portion of the matrix A and B to the second buffers of the
neighbouring eCores without waiting for their computation to finish.  

For blocks of size $32 \times 32$ this is not possible. With each
matrix requiring 4 KBytes of storage, storing the double-buffers for the
operand matrices and the product matrix C would require a total of 20 KBytes of
storage.  However, since the size of the entire code including assembly is 
just over 13 KBytes, this doesn't leave enough space for the double
buffers and the stack.  Hence an alternate buffering scheme was implemented.  

\begin{figure}[ht]
\centering
\includegraphics[width=3.2in]{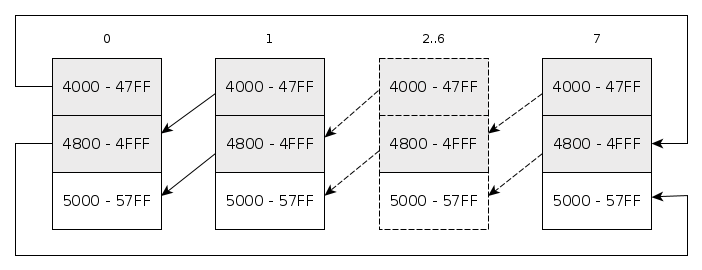}
\caption{Transfer of Matrix A - 1st iteration}
\label{Matrix_A_1}
\end{figure}

\begin{figure}[ht]
\centering
\includegraphics[width=3.5in,height=3.5in,keepaspectratio]{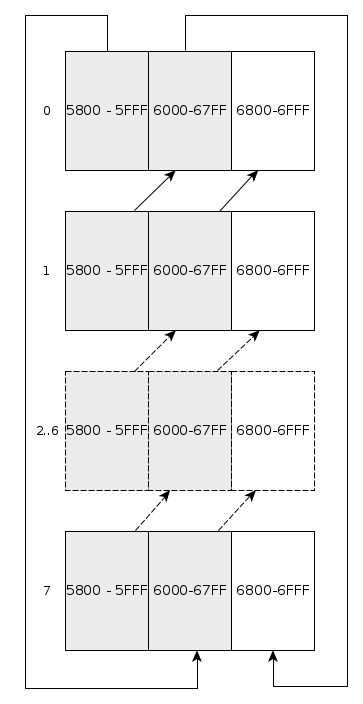}
\caption{Transfer of Matrix B - 1st iteration}
\label{Matrix_B_1}
\end{figure}

In this scheme the matrix A is initially allocated in each eCore from 0x4000 to
0x4FFF and the matrix B from 0x5800 to 0x67FF (4 KBytes each) and the matrix C
is allocated from 0x7000 to 0x7FFF. A buffer of 2 KBytes is allocated adjacent
to each of these matrices, from 0x5000 to 0x57FF for matrix A and 0x6800 to
0x6FFF for matrix B.  Once an eCore is ready to transmit its data, it starts
transferring the lower 2 KBytes of the matrix A onto the buffer for matrix A of
the neighbouring eCore on the left side. This is followed by a transfer of the
lower 2 KBytes of matrix B to the buffer for matrix B of the neighbouring eCore
above it as shown in Figures \ref{Matrix_A_1} and \ref{Matrix_B_1}.

\begin{figure}[ht]
\centering
\includegraphics[width=3.2in]{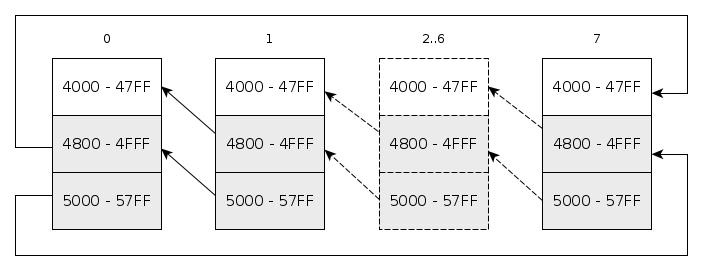}
\caption{Transfer of Matrix A - 2nd iteration}
\label{Matrix_A_2}
\end{figure}

\begin{figure}[ht]
\centering
\includegraphics[width=3.5in,height=3.5in,keepaspectratio]{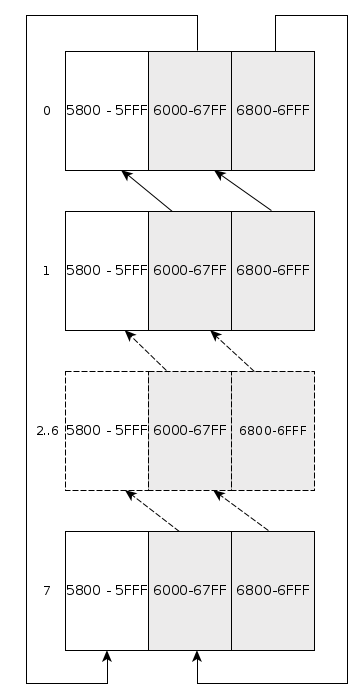}
\caption{Transfer of Matrix B - 2nd iteration}
\label{Matrix_B_2}
\end{figure}

Once all the eCores complete these transfers, they start transferring the upper
halves of the matrices A and B, replacing the lower halves of the corresponding
matrices of the neighbours. The pointers to these two matrices are also changed
accordingly. In the following iteration, communication is performed in the
reverse order as illustrated in Figures \ref{Matrix_A_2} and \ref{Matrix_B_2}.
After changing the pointers to the two matrices again, the allocation of the
matrices would be identical to the initial one.

\subsection*{Off-chip matmul kernel}
\label{matmul_offchip}

For square matrices larger than $256 \times 256$ there is insufficient memory
to perform on-chip matrix multiplication, and it becomes necessary to page
blocks of the matrices from off-chip shared memory. Here we exploit an
analogous algorithm to that used to move blocks of matrices $A$ and
$B$ between eCores for the on-chip case. Namely
blocks of the product matrix $C$ are computed in turn by paging in blocks of
$A$ and $B$ from shared memory. Thus in the $512 \times 512$ case to complete
one $256 \times 256$ block of $C$ requires two $256 \times 256$ blocks of both
$A$ and $B$ to be read from shared memory.

\subsection{Matrix Multiplication Results}
\label{matmul_results}

\subsubsection{Floating Point Performance}

Here, we compare the floating point performance of the matrix multiplication
kernel as a function of sizes of the operand matrices. 

\paragraph{Single-core Floating Point Performance}
The results for a single eCore are shown in Table
\ref{matmul_singlecore_flops}. The maximum size of matrices that are multiplied
is $32 \times 32$ as mentioned earlier. On a single eCore the performance
ranges from 0.85-1.15 GFLOPS or between 70-96\% of peak. 

\begin{table}[ht]
    \begin{center}
        \begin{tabular}{|>{\centering\arraybackslash}p{1.42cm}|>{\centering\arraybackslash}p{1.4cm}|>{\centering\arraybackslash}p{1.5cm}|}
            \hline 
            Matrix Dimensions & GFLOPS & Percentage of Peak (\%) \\
            \hline
            8  $\times$ 8  & 0.85 & 70.5 \\ 
            16 $\times$ 16 & 1.07 & 89.5 \\
            20 $\times$ 20 & 1.11 & 92.5 \\
            24 $\times$ 24 & 1.12 & 93.4 \\
            32 $\times$ 32 & 1.15 & 95.9 \\
            \hline 
        \end{tabular} 
        \caption{Matmul single core floating point performance}
        \label{matmul_singlecore_flops}
    \end{center}
\end{table}

\paragraph{On-chip multi-core Floating Point Performance}

Table \ref{matmul_multicore_flops} shows the floating point performance of the
on-chip multi-core version which was implemented as detailed in Section
\ref{matmul_multicore}. For grid sizes which are able to be fit on the local
memory of the chip (up to $256 \times 256$), the performance is around 85\%
including the data communication between pairs of eCores. (This does not
include the time taken to transfer the initial operand matrices from the
external shared memory to the chip). The table shows the per-core dimensions of
the product matrix C and the number of eCores used to perform the
multiplication.  With a per-core matrix size of $32 \times 32$, the overall
matrix dimensions would be $64 \times 64$ when running on $2 \times 2$ eCores,
$128 \times 128$ on $4 \times 4$ eCores and $256 \times 256$ on $8 \times 8$
eCores.

\begin{table}[ht]
    \begin{center}
        \begin{tabular}{|p{1.2cm}|p{0.87cm}|p{0.6cm}|p{0.87cm}|p{0.6cm}|p{0.87cm}|p{0.6cm}|}
            \hline 
            \multirow{3}{*}{Matrix C} & \multicolumn{6}{c|}{Num of eCores} \\\cline{2-7}
                                      & \multicolumn{2}{c|}{2 $\times$ 2} & \multicolumn{2}{c|}{4 $\times$ 4}  & \multicolumn{2}{c|}{ 8 $\times$ 8} \\\cline{2-7} 
            (per-core) & GFLOPS & \%  & GFLOPS & \%  & GFLOPS & \%  \\
            \hline
            8 $\times$ 8   & 1.25 & 26.1\%   & 5.07  & 26.4\%  &  20.30 & 26.4\%  \\
            16 $\times$ 16 & 3.12 & 65.1\%   & 12.76 & 66.5\%  &  51.41 & 66.9\%  \\
            20 $\times$ 20 & 3.58 & 74.7\%   & 14.36 & 74.8\%  &  57.62 & 75.0\%  \\
            24 $\times$ 24 & 3.84 & 80.1\%   & 15.43 & 80.4\%  &  62.17 & 81.0\%  \\
            32 $\times$ 32 & 4.06 & 84.7\%   & 16.27 & 84.7\%  &  65.32 & 85.1\%  \\
            \hline 
        \end{tabular} 
        \caption{Matmul multi core on-chip floating point performance}
        \label{matmul_multicore_flops}
    \end{center}
\end{table}

The on-chip matrix multiplication of two $256 \times 256$ matrices can be
broken down into the computation of a $32 \times 32$ matrix product by each
eCore and the transfer of the two operand matrices A and B totalling 8 KBytes
to the neighbouring eCore in each iteration. Considering 1.15 GFLOPS for the
matrix product by a single eCore (from Table \ref{matmul_singlecore_flops}) and
2GB/s transfer rate between eCores (from results in Section
\ref{performance_char}), the maximum theoretical performance can be estimated
to be roughly 68 GFLOPS. From Table \ref{matmul_multicore_flops}, the
performance achieved by the code is around 65 GFLOPS which is very close to the
estimate.

\paragraph{Off-chip multi-core Floating Point Performance}

The performance drops for sizes larger than $256 \times 256$ due to the need
for multiple transfers of blocks to and from the shared memory as the algorithm
progresses as discussed earlier. The results are shown in Table
\ref{large_matrix_results}. A per-core matrix size of $32 \times 32$ is used to
perform the multiplication of large matrices of sizes $512 \times 512$ and $1024
\times 1024$. To build the result for the large matrix size $1536 \times
1536$, a per-core size of $24 \times 24$ is used and hence the overall
performance in GFLOPS is a bit worse than the other two cases. In all the
cases, the off-chip memory transfer dominates the overall performance
with around 86-90\% of the total time taken being spent on the block DMA
transfers in and out of shared memory and 10-13\% of the total time taken being
spent on the computation.

\begin{table}[ht]
    \begin{center}
        %\begin{tabular}{|p{1.6cm}|p{0.87cm}|p{0.6cm}|p{1.2cm}|p{1.5cm}|}
        \begin{tabular}{|c|c|p{0.8cm}|p{1.0cm}|p{1.76cm}|}
            \hline 
            Matrix C & GFLOPS & \% of Peak & \% Computation & \% Shared Mem Transfers\\
            \hline
            512 $\times$ 512 & 8.32 & 10.8   \% & 12.8 \% & 87.2 \% \\
            1024 $\times$ 1024 & 8.52 & 11.1 \% & 13.1 \% & 86.9 \% \\
            1536 $\times$ 1536 & 6.34 & 8.2  \% & 10.9 \% & 89.1 \% \\
            \hline 
        \end{tabular} 
        \caption{Floating point performance for larger matrices}
        \label{large_matrix_results}
    \end{center}
\end{table}

To analyse the performance of the off-chip matrix multiplication, we consider
the multiplication of two $512 \times 512$ matrices. Each matrix can be
considered as consisting of four blocks of $256 \times 256$ elements.  Each
iteration in the outer-most loop in the algorithm involves transferring one
block of matrix A and one block of matrix B from the shared memory to the chip
and having all the 64 eCores perform parallel multiplication to produce an
intermediate block result. The transfer of two blocks of $256 \times 256$
elements (512 KBytes) takes around 0.0034 seconds at 150MBytes (from results in
Section \ref{performance_char}). The computation of the block matrix product
takes around 0.00051 seconds at 65.32 GFLOPS (from Table
\ref{matmul_multicore_flops}). Thus the ratio of computation to off-chip
transfers is roughly 1:6.5. From the result in Table \ref{large_matrix_results}
the ratio of computation to off-chip data transfer is 1:6.8 which is very close
to the estimate.

\subsubsection{Weak Scaling}

\begin{figure}[ht]
\centering
\includegraphics[width=3.5in,height=3.5in,keepaspectratio]{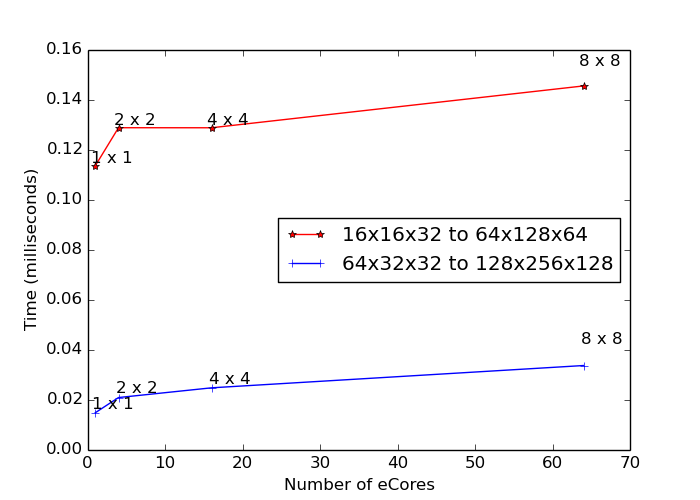}
\caption{Weak Scaling - Number of eCores vs Time}
\label{matmul_weak_scaling}
\end{figure}

In this experiment, the number of eCores is increased from 1 to 64 while
increasing the problem sizes accordingly. Two separate configurations are
tested and the running time for each configuration is plotted in Figure
\ref{matmul_weak_scaling}. The problem sizes are shown as $M \times N \times
K$. Each of these problems are run (wherever possible) on an eCore
configuration of $1 \times 1$, $2 \times 2$, $4 \times 4$ and $8 \times 8$. In
the first configuration, the problem size is increased from $16 \times 16
\times 32$ (1 eCore) to $64 \times 128 \times 64$ ($8 \times 8$ eCores). In the
second configuration, the problem size is increased from $64 \times 32 \times
32$ (1 eCore) to $128 \times 256 \times 128$ ($8 \times 8$ eCores).  The time
taken increases initially due to increasing data communication between the
eCores.  This increase quickly levels out as communication between independent
pairs of eCores is overlapped.

\subsubsection{Strong Scaling}

\begin{figure}[ht]
\centering
\includegraphics[width=3.5in,height=3.5in,keepaspectratio]{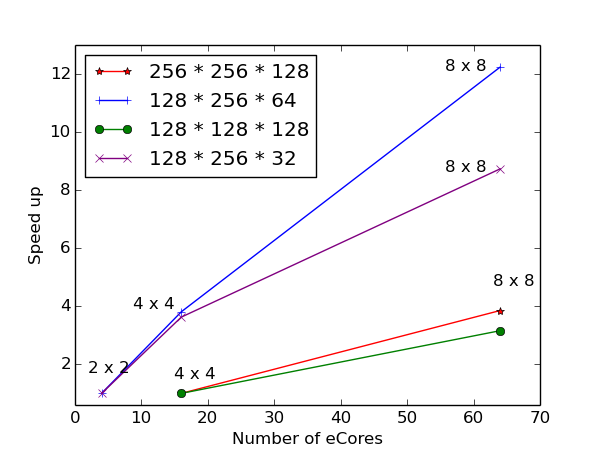}
\caption{Strong Scaling - Number of eCores vs Speedup}
\label{matmul_strong_scaling}
\end{figure}

In this experiment, the number of eCores is increased from 1 to 64 while
keeping the problem size fixed. This is repeated for four different problem
sizes. Each of these problems are run (wherever possible) on an eCore
configuration of $2 \times 2$, $4 \times 4$ and $8 \times 8$. The speed-up
achieved is shown in Figure \ref{matmul_strong_scaling}.  The problem sizes are
shown as $M \times N \times K$. When the number of eCores is quadrupled, a
speed up of close to 4 is achieved. Better results are achieved for larger
problem sizes as expected.

\section{Related Work and Comparison with other systems}
\label{related_work}

Similar many-core coprocessor systems include the Intel 80-core terascale
coprocessor~\cite{mattsonintel}, the 48-core single-chip cloud computer
(SCC)~\cite{mattson201048,vangal200880}, the 64-core Tilera Tile64
SoC~\cite{bell2008tile64,serres2011experiences}, the recent Intel Xeon Phi
accelerator~\cite{heinecke2013design}. Use of low power ARM based SoC
processors~\cite{Mitra2013} and many-core accelerators such as the TI C66X
DSP~\cite{Igual2012Unleashing,stotzer2013openmp} for high performance computing
are also of increasing interest. A comparison of some of these systems is shown
in Table \ref{table:comparison}.

\begin{table}[h]
\begin{center}
\begin{tabular}{|p{1.6cm}|p{1.2cm}|p{1.1cm}|p{1.3cm}|p{1.3cm}|}
\hline
 & \textbf{TI C6678 Multicore DSP} &  \textbf{Tilera 64-core chip} &
 \textbf{Intel 80-core Terascale Processor} & \textbf{Epiphany 64-core
 coprocessor} \\ 
\hline
Chip Power(W) & 10 & 35 & 97 & 2 \\
\hline
Cores & 8 & 64 & 80 & 64 \\
\hline
Max GFLOPS & 160 & 192 & 1366.4 & 76.8 \\
\hline
Clock Speed(GHz) & 1.5 & 0.9 & 4.27 & 0.6\\
\hline
\end{tabular}
\caption{Comparison of Epiphany with other systems}
\label{table:comparison}
\end{center}
\end{table}

The closest related work is that of Mattson et al.~\cite{mattsonintel}. This
paper describes the challenges in programming the Intel 80-core
network-on-a-chip Terascale Processor. With 80 cores running at 4.27 GHz,
Mattson et al. ran the same five-point star-shaped heat diffusion stencil
kernel at 1 single precision TFLOPS while consuming 97 Watts. This is a
significant contribution as it demonstrated that high performance could be
delivered using such many-core coprocessors. It equates to a performance of
just over 10 GFLOPS/watt. By contrast and assuming 2 watts power usage, the
Epiphany system used here achieves roughly 32 GFLOPS/watt. (The actual power
consumed by the chip has not been measured yet, although we are currently
making efforts to profile the actual power consumption.)

In more general work on stencil implementations, Wittman et al.
\cite{wittmann_multicore} explore the explicit use of shared caches in
multicore systems for implementing a pipelined temporal blocking approach for
stencil computation. This technique performs multiple in-cache updates on each
grid point before the result is moved back to memory. Datta et al.
\cite{datta2009} also look at improving stencil computation by examining tiling
techniques that leverage both spatial and temporal blocking of computation in
order to increase data reuse.  Melot et al. \cite{melot2012} address the
performance limitation seen on many-core processors due to the use of off-chip
memory by implementing pipelined streaming computations. This work describes
the advantages of employing an on-chip pipelined merge sort algorithm on the 48
core Intel SCC in an attempt to overcome the limitation of slow off-chip memory
access. 

We extended the work of Sapir\cite{yanivmatmul}, who describes a parallel
matrix multiplication algorithm for matrices which are too large to be fit on
the device memory. This involves block transfers between rows or columns of
nodes using Cannon's algorithm. The transfers are typically between nearest
neighbours and is well suited for 2D mesh architectures such as the Epiphany.
Algorithms such as SUMMA\cite{summa} and PUMMA\cite{pumma} are well known
distributed algorithms for matrix multiplication involving block distribution
of the matrices between processors not necessarily arranged in a grid. SUMMA
also has the advantage of requiring less workspace per node enabling larger
problems to be run.

\section{Conclusion \& Future Work} 
\label{conclusion}

In this paper, we explored the Adapteva Epiphany 64-core Network-on-chip
coprocessor. Different strategies for stencil based application codes running
purely on device memory were implemented. A tuned matrix multiplication kernel
was implemented for multiplying matrices which fit in the device memory along
with a buffering method to overcome the relatively small memory per core.
Using this a building block, an algorithm for multiplying large matrices was
implemented. Micro-benchmarks were used to evaluate the performance of several
basic compute and communication operations on the Epiphany system. The process
of mapping an application kernel to the Epiphany architecture was demonstrated.
It was noted that the low-level C programming primitives available in the
Epiphany SDK provided ease of use in programming the Epiphany.  However,
further work towards implementation of familiar programming models such as
OpenCL~\cite{stoneopencl} and the recently launched OpenMP Accelerator
model~\cite{openmpaccelerator} for the Epiphany is of great interest.
Increasing problem sizes for stencil kernels and implementation of an efficient
pipelined kernel for large problem sizes is also of interest. We aim to extend
our current work by employing a pipelined algorithm for stencil computation
using both spatial and temporal blocking in order to process much higher grid
sizes. Such an algorithm would ensure that computation is performed for a
number of iterations before the data is moved out of the local memory and new
data is brought in.  Multiple iterations of computation can be performed for
those points for which dependent points are in the same block. When no more
iterations are possible, another block would be streamed into the local memory
to ensure that computation can move forward. 

In our experiments, on-chip DMA bandwidth of 2 GBytes/s, off-chip shared memory
access bandwidth of 150 MBytes/s and on-chip memory latency of 11 ns for
nearest neighbour were observed. Floating point performance of roughly 64
GFLOPS (82\% of peak) was achieved for an optimized stencil kernel involving
communication of data and a performance of roughly 65 GFLOPS (85\% of peak) was
achieved for an optimized on-chip parallel matrix multiplication. This
corresponds to a power efficiency of roughly 32 GFLOPS/Watt on the 64-core
chip. However, extracting maximum performance out of the system requires
considerable effort on the part of the programmer at this stage. The relatively
slow external shared memory interface becomes a bottleneck when scaling to
large problem sizes. If these are addressed, with future versions expected to
scale to 4096 cores with a peak floating point performance of 5 TFLOPS and
power efficiency of 70 GFLOPS/Watt, the Epiphany would be a promising platform
for energy efficient high performance computing.

 
\section*{Acknowledgment}    

This work is supported in part by the Australian Research Council Discovery
Project DP0987773. 

% trigger a \newpage just before the given reference number - used to balance
% the columns on the last page adjust value as needed - may need to be
% readjusted if the document is modified later
%\IEEEtriggeratref{8} The "triggered" command can be changed if desired:


% can use a bibliography generated by BibTeX as a .bbl file BibTeX
% documentation can be easily obtained at:
% http://www.ctan.org/tex-archive/biblio/bibtex/contrib/doc/ The IEEEtran
% BibTeX style support page is at:
%\bibliographystyle{IEEEtran} argument is your BibTeX string definitions and
%bibliography database(s) \bibliography{IEEEabrv,../bib/paper}  <OR>
%manually copy in the resultant .bbl file set second argument of \begin to
%the number of references (used to reserve space for the reference number
%labels box)
%\newcommand{\BIBdecl}{\setlength{\itemsep}{0.25 em}}
\def\IEEEbibitemsep{0pt plus .5pt}
\bibliographystyle{IEEEtran} \bibliography{IEEEabrv,./ref} 
\end{document}